\documentclass[aps,prd,onecolumn,groupedaddress,showpacs,nofootinbib,amssymb]{revtex4}
\usepackage[dvips]{graphicx}
\usepackage{amssymb}
\usepackage{amsmath}
\usepackage{graphicx}
\usepackage{amsfonts}
\usepackage{bm}

\begin{document}

\title{A Note on Schwarzschild de Sitter Black Holes in Mimetic $F(R)$ Gravity}

\author{V.K. Oikonomou$^{1,2}$}\,\thanks{v.k.oikonomou1979@gmail.com}
\affiliation{
$^{1)}$ Tomsk State Pedagogical University, 634061 Tomsk \\
$^{2)}$  Laboratory of Theoretical Cosmology, Tomsk State University of Control Systems\\ 
and Radioelectronics (TUSUR), 634050 Tomsk, Russia 
}

\begin{abstract}
In this brief note we investigate the conditions under which a Schwarzschild de Sitter black hole spacetime is a solution of the mimetic $F(R)$ gravity with Lagrange multiplier and potential. As we demonstrate, the resulting mimetic $F(R)$ gravity is a slight modification of the ordinary $F(R)$ gravity case, however the resulting perturbation equations are not in all cases identical to the ordinary $F(R)$ gravity case. In the latter case, the perturbation equations are identical to the ones corresponding to the Reissner-Nordstr\"{o}m anti-de Sitter black hole.
 
\end{abstract}


\maketitle

\section{Introduction}

One of the most appealing features of modified gravity models \cite{fr1,fr2,fr3,fr4,fr5,fr6,fr7,fr8}, is that these can consistently explain the late-time acceleration era which was firstly observed in the late 90's \cite{riess}. In addition to the late-time acceleration, the early-time acceleration era \cite{inflation1,inflation2,inflation3,inflation3a,inflation4,inflation5,inflation6,inflation7} can be incorporated in the same theoretical framework with the late-time acceleration, if these are described by modified gravity models, see for example \cite{sergeinojiri,sergeioikonomou1,sergeioikonomou2,sergeioikonomou3,sergeioikonomou4}. With regards to the late-time acceleration era, the late-time acceleration itself is attributed to a perfect fluid of negative effective pressure, called dark energy, and this dark energy overwhelms over all other perfect fluids, in the present day total energy density, having a percentage of ($\sim 68.3\%$). For an important stream of papers on dark energy we refer the reader to \cite{de1,de2,de3,de4,de6,de7} and references therein. The rest of the energy density of our Universe is described by ordinary luminous matter ($\Omega_m\sim4.9\%$), and of dark matter ($\Omega_{DM}\sim 26.8\%$). With regards to dark matter, a lot of proposals exist aiming to the successful description of this component of the Universe content, and in some cases it is described by a particle which weakly interacts with ordinary matter, see for example \cite{oikonomouvergados} for some proposals. Another quite appealing description of dark matter is provided by mimetic gravitational theories of any kind, firstly introduced in \cite{mukh1}, in the context of Einstein-Hilbert gravity, and later developed in \cite{mukh2,golovnev,NO2,mim1,mim3,mim4,mim5,mim6,mim7,mim8,mim9,mim10,mim11,mim12,mim13,mim14,mim15,mim16,mim17}.

In this paper we apply the analysis of \cite{mim14} in the case of the mimetic $F(R)$ gravity Schwarzschild de Sitter black holes, in which case we use again the Lagrange multiplier formalism \cite{lagra1,lagra2} in order to realize the mimetic constraint. For a recent work on the Schwarzschild de Sitter black holes in $F(R)$ gravity, see \cite{salvatorerecent}. As we demonstrated in \cite{mim15}, the Reissner-Nordstr\"{o}m-anti de Sitter black holes are solutions of a trivial mimetic $F(R)$ gravity, and the resulting equations of motion, when these are linearly perturbed, they lead to anti-evaporation phenomena, like in \cite{rnsergei1} for ordinary $F(R)$ gravity. As we demonstrated in \cite{mim15}, the resulting mimetic $F(R)$ gravity was the same as an ordinary $F(R)$ with an additional cosmological constant, and therefore the linearly perturbed equations of motion lead to anti-evaporation phenomena \cite{rnsergei1,rnsergei2,hawking,sergeiearly1,antievapsergei7}. For some relevant works on black holes in modified gravity see \cite{bh1,bh2,bh3,bh4,bh5,bh6,bh7,bh8,bh9,bh10,bh11,bh12,bh14} and references therein. As we demonstrate in the present work, the Schwarzschild de Sitter black hole is a solution of trivial mimetic $F(R)$ gravity, which is an ordinary $F(R)$ gravity with a cosmological constant. Also, the perturbed equations of motion can be identical to the ones corresponding to the Reissner-Nordstr\"{o}m anti-de Sitter black holes, leading to anti-evaporation. Also there are cases that the perturbed equations of motion have different form, from the ones corresponding to Reissner-Nordstr\"{o}m anti-de Sitter black holes, in which case anti-evaporation also occurs, as in Ref. \cite{rnsergei2}.

This letter is organized as follows: In section II, we present the theoretical framework of mimetic vacuum $F(R)$ gravity with Lagrange multiplier and mimetic potential. We also investigate when the Schwarzschild de Sitter black hole is a solution of the mimetic $F(R)$ gravity in the absence of any matter fluid. In section III we investigate the linear perturbations of the Schwarzschild de Sitter black hole equations of motion and discuss all the possibilities.  The conclusions follow in the end of the paper.

\section{Mimetic $F(R)$ Gravity and Schwarzschild de Sitter Black Holes}

Up to now there exist several works on mimetic $F(R)$ gravity \cite{NO2,mim14,mim15,mim16,mim17}, and this theory was firstly developed in \cite{NO2}, by using the Lagrange multiplier formalism \cite{lagra1,lagra2}. In the mimetic gravity formalism, the physical metric is written in terms of an auxiliary scalar field $\phi$ and an auxiliary metric tensor $\hat{g}_{\mu \nu}$, as follows,
\begin{equation}\label{metrpar}
g_{\mu \nu}=-\hat{g}^{\mu \nu}\partial_{\rho}\phi \partial_{\sigma}\phi
\hat{g}_{\mu \nu}\, .
\end{equation}
Practically, in the mimetic gravity formalism we make use of the internal conformal degrees of freedom of the metric, so by varying the action of mimetic Einstein-Hilbert gravity, with respect to the metric $\hat{g}_{\mu \nu}$ we obtain the following constraint, 
\begin{equation}\label{impl1}
g^{\mu \nu}(\hat{g}_{\mu \nu},\phi)\partial_{\mu}\phi\partial_{\nu}\phi=-1\,
,
\end{equation}
to which we shall refer to as the ``mimetic constraint''. Note that under the transformation $\hat{g}_{\mu \nu}=e^{\sigma (x)}g_{\mu \nu}$, the Eq. (\ref{metrpar}) is invariant and effectively the auxiliary metric is absent from the final action. In the context of mimetic $F(R)$ gravity with Lagrange multiplier $\lambda$ and potential $V(\phi)$ without matter fluids, the action is equal to \cite{NO2},
\begin{equation}\label{actionmimeticfraction}
S=\int \mathrm{d}x^4\sqrt{-g}\left ( F\left(R(g_{\mu \nu})\right
)-V(\phi)+\lambda \left(g^{\mu \nu}\partial_{\mu}\phi\partial_{\nu}\phi
+1\right)\right )\, .
\end{equation}
Upon varying the action of Eq. (\ref{actionmimeticfraction}), with respect to the metric $g_{\mu \nu}$, we acquire the following equations of motion,
\begin{align}\label{aeden}
& \frac{1}{2}g_{\mu \nu}F(R)-R_{\mu
\nu}F'(R)+\nabla_{\mu}\nabla_{\nu}F'(R)-g_{\mu \nu}\square F'(R)\\ \notag &
+\frac{1}{2}g_{\mu \nu}\left (-V(\phi)+\lambda \left( g^{\rho
\sigma}\partial_{\rho}\phi\partial_{\sigma}\phi+1\right) \right )-\lambda
\partial_{\mu}\phi \partial_{\nu}\phi =0 \, .
\end{align}
In addition, the variation of the action (\ref{actionmimeticfraction}) with respect to the scalar field $\phi$, yields,
\begin{equation}\label{scalvar}
-2\nabla^{\mu} (\lambda \partial_{\mu}\phi)-\frac{\mathrm{d}V(\phi)}{\mathrm{d}\phi}=0\, ,
\end{equation}
while the variation with respect to the Lagrange multiplier $\lambda$, yields the unimodular constraint of Eq. (\ref{impl1}), as was expected, that is,
\begin{equation}\label{lambdavar}
g^{\rho \sigma}\partial_{\rho}\phi\partial_{\sigma}\phi=-1\, ,
\end{equation}

Having presented the essential features of mimetic $F(R)$ gravity with potential and Lagrange multiplier, now we investigate under which conditions the mimetic $F(R)$ gravity has the Schwarzschild de Sitter black hole spacetime as a solution of the corresponding field equations. 


The question is what is the form of the spherically symmetric metric which is a solution of the mimetic $F(R)$ gravity theory, with the spacetime curvature being constant. Consider the following spherically symmetric metric, 
\begin{equation}\label{metricstatic}
\mathrm{d}s^2=g_{\mu \nu }\mathrm{d}x^{\mu }\mathrm{d}x^{\nu }=-A(r)\mathrm{d}t^2+B(r)\mathrm{d}r^2+r^2\mathrm{d}\Omega^2\, ,
\end{equation}
where $A(r)$ and $B(r)$ are smooth and differentiable functions of $r$, and in moreover, $\mathrm{d}\Omega^2$ denotes the metric of unit 2-sphere. Consider the class of metrics which satisfy the following conditions,
\begin{equation}\label{constraints}
A(r)=\frac{1}{B(r)},\,\,\,R=R_0\, ,
\end{equation}
with $R$ denoting as usual the Ricci scalar, and $R_0$ a positive constant. The Ricci scalar corresponding to the metric (\ref{metricstatic}), reads,
\begin{equation}\label{ricciscalar}
R=-A''(r)-\frac{4}{r}A'(r)-\frac{2}{r^2}A(r)+\frac{2}{r^2}\, ,
\end{equation}
with the ``prime'' denoting differentiation with respect to the radial coordinate $r$. For $R=R_0$, we get,
\begin{equation}\label{diffeqn}
-A''(r)-\frac{4}{r}A'(r)-\frac{2}{r^2}A(r)+\frac{2}{r^2}=R_0\, ,
\end{equation}
the solution of which is,
\begin{equation}\label{solvesol}
A(r)=1-\frac{r^2}{12}R_0+\frac{C_1}{r}+\frac{C_2}{r^2}\, .
\end{equation}
By setting the arbitrary integration constants as $C_1=-M$ and $C_2=Q$, we obtain,
\begin{equation}\label{solveeqn1}
A(r)=1-\frac{R_0\,r^2}{12}-\frac{M}{r}+\frac{Q}{r^2}\, .
\end{equation}
When $Q=0$, the metric becomes,
\begin{equation}\label{metricressin}
\mathrm{d}s^2=-\Big{(}1-\frac{R_0\,r^2}{12}-\frac{M}{r}\Big{)}\mathrm{d}t^2+\frac{1}{\Big{(}1-\frac{R_0\,r^2}{12}-\frac{M}{r}\Big{)}}\mathrm{d}r^2+r^2\mathrm{d}\Omega^2\, ,
\end{equation}
which describes the Schwarzschild-de Sitter black hole spacetime. This spacetime has two event horizons, a cosmological and an event horizon, which can be found by solving the equation $\frac{1}{g_{rr}}=0$. We shall be interested in the limiting case for which the two horizon coincide. The it can be shown that the metric becomes identical to that of the Nariai spacetime \cite{rnsergei2},
\begin{equation}\label{metricperturbed}
\mathrm{d}s^2=\frac{1}{M_{+}^2\cosh^2 x}\left(\mathrm{d}\bar{t}^2-\mathrm{d}x^2 \right)+\frac{1}{M_{+}^2}\mathrm{d}\Omega^2\, ,
\end{equation}
where $x$ is a variable defined in terms of the radial coordinate $r$, and $M$ is a mass scale. For details on the derivation of the Nariai metric from the Schwarzschild-de Sitter black hole spacetime, we refer to Refs. \cite{rnsergei1,rnsergei2,mim14}. It is exactly the Nariai metric of Eq. (\ref{metricperturbed}) which we will perturb, but before going to this, we perform an analysis of the mimetic $F(R)$ solutions. As we demonstrate, the resulting mimetic $F(R)$ theory is a trivial extension of $F(R)$ gravity.

Indeed, by using Eqs. (\ref{scalvar}) and (\ref{lambdavar}), the equations of motion of Eq. (\ref{aeden}), take the following form,
\begin{align}\label{equationofmotionbasic}
& \frac{1}{2}g_{\mu \nu}F(R)-R_{\mu
\nu}F'(R)+\nabla_{\mu}\nabla_{\nu}F'(R)-g_{\mu \nu}\square F'(R)+
\frac{1}{2}g_{\mu \nu}\left (-V(\phi)\right )-\lambda
\partial_{\mu}\phi \partial_{\nu}\phi =0 \, ,
\end{align}
and since $R=R_0$ and therefore a constant, we obtain,
\begin{align}\label{equationofmotionbasicconstantcurv}
& \frac{1}{2}g_{\mu \nu}F(R)-R_{\mu
\nu}F'(R)-
\frac{1}{2}g_{\mu \nu}V(\phi)-\lambda
\partial_{\mu}\phi \partial_{\nu}\phi =0 \, ,
\end{align}
Upon contraction, Eq. (\ref{equationofmotionbasicconstantcurv}) becomes,
\begin{equation}\label{contactedeqn}
2 F(R)-RF'(R)-2V(\phi )-\lambda g^{\mu \nu}\partial_{\mu}\phi \partial_{\nu}\phi=0\, ,
\end{equation}
and by also taking into account Eq. (\ref{lambdavar}), we obtain,
\begin{equation}\label{finalfreqn}
F(R)=\frac{RF'(R)}{2}+V(\phi )-\frac{\lambda }{2}\, .
\end{equation}
By combining Eqs. (\ref{finalfreqn}) and (\ref{equationofmotionbasicconstantcurv}), we obtain,
\begin{equation}\label{finaleqn}
\left (  \frac{g_{\mu \nu}R}{4}- R_{\mu
\nu}  \right )F'(R)-g_{\mu \nu }\left(\frac{3\lambda }{4}\right)=0\, .
\end{equation}
The term $\left (  \frac{g_{\mu \nu}R}{4}- R_{\mu
\nu}  \right )$ for the Schwarzschild de Sitter metric of Eq. (\ref{metricressin}) is identical to zero, so the only solution for the mimetic $F(R)$ gravity is $V(\phi)=\Lambda$, and $\lambda=0$, with $\Lambda$ being a constant. Hence, the resulting mimetic theory is a trivial extension of the $F(R)$ gravity, at least when a constant curvature spherically symmetric solution is considered. Note that in all cases, we can either have $F'(R)=0$ or $F'(R_0)\neq 0$, for the Schwarzschild de Sitter case, which will modify the resulting anti-evaporation picture in comparison to the ordinary $F(R)$ gravity case, studied in Ref. \cite{rnsergei2}. In the following we shall consider the implications of these considerations in detail, by investigating the linear perturbations of the equations of motion. for the metric (\ref{metricperturbed}).

\section{Perturbations of the Mimetic $F(R)$ Schwarzschild de Sitter Black Hole}

Having described in brief the conditions under which the Schwarzschild de Sitter black hole is a solution of mimetic $F(R)$ gravity, we now proceed to the perturbations of the field equations. Recall that the mimetic $F(R)$ gravity solution for $R=R_0$ has the following trivial form, 
\begin{equation}\label{constrscenarioi}
V(\phi )=\Lambda,\,\,\, \lambda =0,\,\,\, F(R_0)=\frac{RF'(R_0)}{2}+\Lambda\, .
\end{equation}
Obviously the resulting picture is slightly different from the ordinary $F(R)$ gravity, due to the appearance of the constant term $\Lambda$, but the constant  $\Lambda$ can be absorbed into the redefinition of the $F(R)$ gravity, so this is why it is slightly different from the ordinary $F(R)$ case. It is owing to this trivial difference, that the perturbations around the constant curvature solution corresponding to the mimetic and ordinary $F(R)$ are almost identical. Note that $F'(R_0)$ and $\Lambda$ are not constrained, so the possible choices are that these can be positive or negative, zero or non-zero. This will play some role in the following analysis. The resulting equations of motion on which we perform perturbations are,
\begin{align}\label{equationsmotionfinalscenarioi}
& \frac{1}{2}g_{\mu \nu}F(R)-R_{\mu
\nu}F'(R)-
\frac{1}{2}g_{\mu \nu}\Lambda =0 \, .
\end{align}
In addition, we perturb the resulting Nariai metric of Eq. (\ref{metricperturbed}), as follows,
\begin{equation}\label{metricperturbed1}
\mathrm{d}s^2=\frac{e^{2\rho(x,\bar{t})}}{M_{+}^2\cosh^2 x}\left(\mathrm{d}\bar{t}^2-\mathrm{d}x^2 \right)+\frac{e^{-2\varphi(x,\bar{t})}}{M_{-}^2}\mathrm{d}\Omega^2\, ,
\end{equation}
where the functions $\rho (x,\bar{t})$ and $\varphi (x,\bar{t})$ are smooth and differentiable functions of $x$ and $\bar{t}$, defined as follows,
\begin{equation}\label{functionsperturbations}
\rho (x,\bar{t})=-\ln (\cosh x)+\delta \rho,\,\,\, \varphi=\delta \varphi\, .
\end{equation} 
By using the perturbed metric (\ref{metricperturbed1}) and, we obtain the following set of equations corresponding to the  $(\bar{t},\bar{t})$, $(x,x)$, $(x,\bar{t})$, $(\bar{t},x)$, $(\theta,\theta)$ and to $(\phi ,\phi )$ components of the metric,
\begin{align}\label{comp1}
&\frac{e^{2\rho}}{2M^2}F(R)-\Big{(}\ddot{\rho}+2\ddot{\varphi}+\frac{\partial^2\rho}{\partial x^2}-2\dot{\varphi}^2-\frac{\partial\rho}{\partial x}\frac{\partial \varphi}{\partial x }-2\dot{\rho}\dot{\varphi}\Big{)}F'(R)+\frac{\partial^2F'(R)}{\partial \bar{t}^2}-\dot{\rho}\frac{\partial F'(R)}{\partial \bar{t}}-\frac{\partial \rho}{\partial x}\frac{\partial F'(R)}{\partial x}\\ \notag &+e^{2\varphi }\frac{\partial }{\partial x }\Big{(}e^{-2\varphi }\frac{\partial F'(R)}{\partial x}\Big{)}-e^{2\varphi }\frac{\partial }{\partial \bar{t} }\Big{(}e^{-2\varphi }\frac{\partial F'(R)}{\partial \bar{t}}\Big{)}-\frac{e^{2\rho}}{2M^2}\Lambda=0\, ,
\end{align}
\begin{align}\label{comp2}
&-\frac{e^{2\rho}}{2M^2}F(R)-\Big{(}\ddot{\rho}+2\frac{\partial^2\varphi }{\partial x^2}-\frac{\partial^2\rho}{\partial x^2}-2\left(\frac{\partial\rho}{\partial x}\right)^2-2\frac{\partial\rho}{\partial x}\frac{\partial\varphi}{\partial x}-2\dot{\rho}\dot{\varphi}\Big{)}F'(R)+\frac{\partial^2F'(R)}{\partial x^2}-\dot{\rho}\frac{\partial F'(R)}{\partial \bar{t}}-\frac{\partial \rho}{\partial x}\frac{\partial F'(R)}{\partial x}\\ \notag &-e^{2\varphi }\frac{\partial }{\partial x }\Big{(}e^{-2\varphi }\frac{\partial F'(R)}{\partial x}\Big{)}-e^{2\varphi }\frac{\partial }{\partial \bar{t} }\Big{(}e^{-2\varphi }\frac{\partial F'(R)}{\partial \bar{t}}\Big{)}+\frac{e^{2\rho}}{2M^2}\Lambda=0\, ,
\end{align}
\begin{align}\label{comp3}
& -\left( -2\frac{\partial \dot{\varphi} }{\partial x }-2 \frac{\partial \varphi }{\partial x }\dot{\varphi}-2 \frac{\partial \rho }{\partial x }\dot{\varphi}-2\dot{\rho}\frac{\partial \varphi}{\partial x }\right)F'(R)+\frac{\partial^2F'(R) }{\partial x \partial \bar{t} }-\dot{\rho}\frac{\partial F'(R)}{\partial x } -\frac{\partial \rho }{\partial x }\frac{\partial F'(R)}{\partial \bar{t} }=0\, ,
\end{align}
\begin{align}\label{comp4}
& \frac{e^{-2\varphi}}{2M^2}F(R)-\frac{M^2}{M^2}e^{-2(\rho+\varphi)}\left ( -\ddot{\varphi}+2 \frac{\partial^2 \varphi }{\partial x^2}-2\left(\frac{\partial \varphi}{\partial x}\right)^2+2\dot{\varphi}^2\right) F'(R)+F'(R)
\\ \notag & +\frac{M^2}{M^2}e^{-2(\rho+\varphi)}\left ( \dot{\varphi}\frac{\partial F'(R) }{\partial \bar{t}}-\frac{\partial \varphi}{\partial x}\frac{\partial F'(R)}{\partial x}\right)-\frac{M^2}{M^2}e^{-2\rho}\Big{[}-\frac{\partial }{\partial \bar{t}}\left(e^{-2\varphi }\frac{\partial F'(R)}{\partial \bar{t} }\right)+ \frac{\partial }{\partial x}\left(e^{-2\varphi }\frac{\partial F'(R)}{\partial x }\right)\Big{]}-\frac{e^{-2\varphi}}{2M^2}\Lambda\, .
\end{align}
Considering now that the perturbations are located around the constant curvature solution, we can obtain the perturbed equations around $R=R_0$. At this point however it is crucial to present in brief how each term is transformed and also discriminate the case for which $F'(R_0)=0$ or $F'(R_0)\neq 0$. We start off with the transformation properties of each perturbed term, so these are as follows, 
\begin{equation}\label{var1}
\delta \left[ \frac{e^{2\rho}}{2M^2}F(R)\right]=\frac{\delta \rho e^{2\rho}}{2M^2}F(R_0)+\frac{e^{2\rho}}{2M^2}F'(R_0)\delta R\, ,
\end{equation}
\begin{equation}\label{dfedd}
\delta \Big{(}-2\dot{\varphi}^2-\frac{\partial\rho}{\partial x}\frac{\partial \varphi}{\partial x }-2\dot{\rho}\dot{\varphi}\Big{)}F'(R))\simeq -\frac{1}{\cosh^2 x}F''(R_0)\delta R+ \Big{(}-\delta\ddot{\rho}+2\delta\varphi''-2\delta\rho''+2 \tanh x \delta \varphi '\Big{)}F'(R_0)\, .
\end{equation}
\begin{align}\label{firstorederannalyticvariations}
& \delta \left[ \frac{\partial^2F'(R)}{\partial \bar{t}^2}\right ]=F''(R_0)\delta\ddot{ R}\, ,\\ \notag &
\delta \left[ -\dot{\rho}\frac{\partial F'(R)}{\partial \bar{t}}\right ]=-\dot{\rho}F''(R_0)\delta \dot{\rho}\, ,\\ \notag &
\delta \left[ -\frac{\partial \rho}{\partial x}\frac{\partial F'(R)}{\partial x}\right ]=\tanh x F''(R_0) \frac{\partial\delta R}{\partial x }\, ,\\ \notag &
\delta \left[ e^{2\varphi }\frac{\partial }{\partial x }\Big{(}e^{-2\varphi }\frac{\partial F'(R)}{\partial x}\Big{)}\right ]=-F''(R_0)\delta \ddot{R}\, , \\ \notag &
\delta \left[ -e^{2\varphi }\frac{\partial }{\partial \bar{t} }\Big{(}e^{-2\varphi }\frac{\partial F'(R)}{\partial \bar{t}}\Big{)}\right ]=F''(R_0) \frac{\partial^2 \delta R}{\partial x^2 }\, , \\ \notag &
\delta \left[ -\frac{e^{2\rho}}{2M^2}\Lambda\right]=-\frac{e^{2\rho}}{2M^2}\Lambda\delta \rho\, ,
\end{align}
where we also used,
\begin{equation}\label{eauxil}
\frac{\partial^2 \rho}{\partial x^2}=\frac{1}{\cosh ^2 x}+\frac{\partial^2 \delta \rho}{\partial x^2}\, .
\end{equation}
and also only the first order terms in the perturbed variables are taken into account. The Ricci scalar variation at first order reads, 
\begin{equation}\label{olonsygkrotei}
\delta R=- M^2\delta \rho+4M^2\delta \varphi-M^2\cosh ^2 x\left( 2\left(\delta\ddot{\rho }-\frac{\partial^2\delta \rho}{\partial x^2} \right)-4\left(\delta \ddot{\varphi}-\frac{\partial^2\delta \varphi}{\partial x^2} \right)\right)\, .
\end{equation}
At this point we need to study the cases $F'(R_0)\neq 0$ and $F'(R_0)=0$ differently. We start of with the case $F'(R_0)\neq 0$ first. 

\subsubsection{The Case $F'(R_0)\neq 0$}

In the case $F'(R_0)\neq 0$, by taking into account the transformation properties above, and finally Eq. (\ref{constrscenarioi}), the resulting equations of motion become,
\begin{equation}\label{resultingeqns1}
-\frac{\mathrm{d}}{\mathrm{d}x}\left ( -2\delta \varphi+\frac{F''(R_0)}{F'(R_0)}\right)+\tanh x \left( \frac{F''(R_0)}{F'(R_0)}\delta R-2\delta \varphi\right)=0\, ,
\end{equation}
\begin{equation}\label{res1}
\frac{-F'(R_0)+2M^2F''(R_0)}{2M^2}\delta R-F'(R_0)\left (\delta \rho-\delta \varphi-\frac{F''(R_0)}{F'(R_0)}\delta R \right)=0
\end{equation}
\begin{equation}\label{res2}
\frac{2M^2}{\cosh^2 x}\delta \varphi+\delta \rho''-\delta \ddot{\rho}-\frac{F''(R_0)}{2F'(R_0)}\delta \ddot{R}+\frac{F''(R_0)}{2F'(R_0)}\delta R''=0\,. 
\end{equation}
Therefore, the resulting picture of the perturbations of the Schwarzschild de Sitter mimetic $F(R)$ black holes are identical to the ordinary $F(R)$ case perturbations \cite{rnsergei2}, when $F'(R_0)\neq 0$. However, as we demonstrate in the next section, the perturbations of the mimetic $F(R)$ case can be different from the ordinary $F(R)$ case, if $F(R_0)=\Lambda$.

\subsection{The Case $F'(R_0)= 0$ or Equivalently $F(R_0)=\Lambda$}

The mimetic $F(R)$ gravity framework for Schwarzschild de Sitter black hole, differs from the corresponding ordinary $F(R)$ case, due to the fact that there is much more freedom in choosing $F'(R_0)$. In fact in the ordinary $F(R)$ case, $F'(R_0)$ cannot be set equal to zero, since this would fix the function $F(R_0)$ too, with the latter being equal to zero. In fact, in the ordinary $F(R)$ case, the constraint equation between $F(R)$ and $F'(R)$ at $R=R_0$ is,
\begin{equation}\label{ordfrconstr}
F(R_0)=\frac{R}{2}F'(R_0)\, ,
\end{equation}
which can be compared with the constraint of Eq. (\ref{constrscenarioi}). As it can be seen from Eq. (\ref{ordfrconstr}), if $F'(R_0)=0$, this leads to the constraint $F(R_0)=0$ too, but in the mimetic $F(R)$ case, if $F'(R_0)=0$, this leads to the less restrictive constraint $F(R_0)=\Lambda$. Obviously, the $F(R_0)=0$ constraint is more restrictive, since the functional form of the $F(R)$ gravity has to be of the form $F(R)\sim (R-R_0)^n$, with $n$ some positive real number. Clearly in the mimetic case, the functional form of the $F(R)$ gravity can be freely chosen, since any function will obey $F(R_0)=\Lambda$, for a constant curvature $R_0$. As we already mentioned, when $F(R_0)=\Lambda$, this implies that $F'(R_0)=0$, so let us investigate how the perturbations will behave with this constraint. By using the transformations (\ref{var1}), (\ref{dfedd})
and (\ref{firstorederannalyticvariations}), with $F'(R_0)=0$ and $F(R_0)=\Lambda$, we obtain the following equations obeyed by the perturbations of the Nariai metric (\ref{metricperturbed1}),
\begin{equation}\label{perturbedeq1}
F''(R_0)\Big{[} -\frac{1}{\cosh^x}\delta R+\tanh x\frac{\partial \delta R}{\partial x } +\frac{\partial^2 \delta R}{\partial x^2 }\Big{]}=0\, ,
\end{equation}
\begin{equation}\label{perturbedeq2}
F''(R_0)\Big{[} \frac{1}{\cosh^x}\delta R+\delta\ddot{ R} +\tanh x\frac{\partial \delta R}{\partial x }\Big{]}=0\, ,
\end{equation}
\begin{equation}\label{perturbedeq3}
F''(R_0)\Big{[} \frac{\partial \delta \dot{R}}{\partial x } +\tanh x\delta \dot{R}\Big{]}=0\, ,
\end{equation}
\begin{equation}\label{perturbedeq4}
F''(R_0)\Big{[} \delta R+\cosh^2 x\left(-\delta \ddot{R} +\frac{\partial^2 \delta R}{\partial x^2 }\right)\Big{]}=0\, .
\end{equation}
which intriguingly enough are identical to the perturbation equations of mimetic $F(R)$ gravity for the Reissner-Nordstr\"{o}m anti de Sitter black hole, which is clearly different from the Schwarzschild de Sitter black hole, due to the fact that $Q$ in Eq. (\ref{solveeqn1}) is non-zero in the Reissner-Nordstr\"{o}m case. It is expected that the anti-evaporation possibility in this case will be the same in the too cases, but we refer from going into details on this, since a thorough analysis was performed in \cite{rnsergei1,rnsergei2,mim14}.

\section{Conclusions}

In this paper we investigated the conditions under which the Schwarzschild de Sitter black hole is a solution of mimetic $F(R)$ gravity with Lagrange multiplier and mimetic potential. After assuming that the curvature is constant, the resulting mimetic $F(R)$ gravity is a slight modification of the ordinary $F(R)$ case, and particularly it is ordinary $F(R)$ plus a cosmological constant. The resulting perturbation equations in the case $F(R_0)\neq \Lambda$, where $\Lambda$ is the aforementioned cosmological constant, are identical to the ones corresponding to ordinary $F(R)$ case. Intriguingly enough, when $F(R_0)=\Lambda$, the resulting perturbation equations are identical to the ones corresponding to the Reissner-Nordstr\"{o}m anti-de Sitter case, with the difference being that in the case at hand, the $F(R)$ gravity is less restricted in comparison to the Reissner-Nordstr\"{o}m anti-de Sitter case. Hence, although the mimetic $F(R)$ case is a slight modification of the ordinary $F(R)$, we can see that the resulting perturbation equations of the Schwarzschild de Sitter and Reissner-Nordstr\"{o}m anti-de Sitter case are connected.

\section*{Acknowledgments}

This work is supported by Min. of Education and Science of Russia (V.K.O).

\end{document}